\documentstyle[eqsecnum,aps,epsfig]{revtex}

\title{Exact spinor-scalar bound states in a QFT with scalar interactions}

\author{ Volodymyr Shpytko and Jurij Darewych \\
Department of Physics and Astronomy, York University\\ 
Toronto, ON   M3J 1P3  Canada}
\vss

\newcommand{\re}[1]{(\ref{#1})}

\def\xp{x^\prime}

\newcommand{\bl}{\mbox{\boldmath$l$}}
\newcommand{\bj}{\mbox{\boldmath$j$}}
\newcommand{\bJ}{\mbox{\boldmath$J$}}
\newcommand{\bs}{\mbox{\boldmath$s$}}
\newcommand{\bsi}{\mbox{\boldmath$\sigma$}}

\newcommand{\bnab}{\mbox{\boldmath$\bigtriangledown $}}
\newcommand{\bP}{\mbox{\boldmath$P$}}
\newcommand{\bp}{\mbox{\boldmath$p$}}
\newcommand{\br}{\mbox{\boldmath$r$}}
\newcommand{\bx}{\mbox{\boldmath$x$}}
\newcommand{\by}{\mbox{\boldmath$y$}}
\newcommand{\bxp}{\mbox{\boldmath$x^{\prime}$}}

\def\12{\frac12}

\def\ds{\displaystyle}
\def\xp{x^\prime}

\newcommand{\bra}{{\langle}}
\newcommand{\ket}{{\rangle}}
\def\nn{{\nonumber}}

\def\phid{{\phi^\dag }}

\def\di{{\partial}}
\def\ni{{\noindent}}
\def\phib{{\overline \phi}}
\def\Psib{{\overline \Psi}}

\def\phid{{\phi^\dag}}
\def\Psid{{\Psi^\dag}}

\def\vss{{\vskip .5cm}}

\begin{document}

\maketitle

\begin{abstract}
We study two-particle systems in a model quantum field theory, in which  scalar 
particles and spinor particles  
 interact via a mediating scalar field. The Lagrangian of the model is reformulated
by using  covariant Green's functions  to solve for the mediating field in terms of the 
particle fields.  This results in a Hamiltonian in which the mediating-field propagator 
appears directly in the interaction term.
It is shown that exact two-particle eigenstates of the Hamiltonian can be determined.
The resulting relativistic fermion-boson equation is shown to have Dirac and Klein-Gordon
 one-particle limits. Analytic solutions for the bound state energy spectrum are obtained for the 
case of massless mediating fields. 

\end{abstract}

\section{Introduction}

\setcounter{equation}{0}
\renewcommand{\theequation}{1.\arabic{equation}}

\par
There are few examples of analytically solvable relativistic two-body wave equations,
 particularly equations derived from quantum field theory(QFT). The principal exception is the
case of spinless (scalar) particles interacting via a mediating scalar field, the so-called 
scalar Yukawa or Wick-Cutkosky model.  This model, therefore, has served as a 
favourite testing ground for methods of solving bound-state problems in QFT, beginning with 
the Wick \cite{WC1} and Cutkosky \cite{WC2} solution of the Bethe-Salpeter equation in ladder
 approximation.

In  earlier papers \cite{Dar98,BD98} it was shown  that
 the scalar Yukawa model  can be recast in a form such 
that {\sl exact} two-body eigenstates of the Hamiltonian, in the 
canonic equal-time formalism, can be determined for the case where there 
are no free (physical) quanta of the mediating ``chion" field ({\sl i.e.} only virtual 
chions).  This is achieved by the partial elimination of the mediating chion
field by means of Green functions,  by the use of the 
Feshbach-Villars formulation for the scalar particle fields,  and by the use of an ``empty'' 
vacuum state. 
The resulting two-particle bound state mass spectrum, for the case of 
massless chion exchange, was found to be
\begin{equation}\label{1.20}
E_{\pm} = \sqrt{m_1^2 + m_2^2 \pm 2 m_1 m_2 \sqrt{1- {\alpha^2 \over n^2}} } ,
\end{equation}                
where $\alpha$ is the effective dimensionless coupling constant, and $n=1,2,...$ is the principal 
quantum number.  

The expression \re{1.20} for $E(\alpha/n)$  has the shape of a distorted 
semicircle. Indeed, in terms of the variables $x = (E^2 - m_1^2 -m_2^2)/2m_1m_2$
and $y= \alpha / n$, \re{1.20} corresponds to  half of the circle $x^2+y^2=1$.
The upper branch $E_+$ of the energy eigenvalue \re {1.20} starts at 
$m_1+ m_2$ when $\alpha$ is zero,
and decreases monotonically to the value $E_c = \sqrt{m_1^2+m_2^2}$ when 
$\alpha $ reaches the critical value $\alpha_c = n$, beyond which the energy ceases to be 
real.                       
This behaviour is reminiscent of what happens for the one-body Klein-Gordon-Coulomb 
and Dirac-Coulomb systems. From $\ds E_+ =   m_1+m_2 - {1 \over 2} { {m_1 m_2} \over 
{m_1+m_2} } {\alpha^2 \over n^2}  + O(\alpha^4)$ it is clear        
that $E_+ (\alpha)$ has the correct non-relativistic Schr\"odinger (Balmer) limit. It also has the 
expected Klein-Gordon  one-body limit (with scalar Coulombic potential), namely 
$E_{\pm} - m_1 \to \pm m_2 \, \sqrt{1- \alpha^2/n^2}$ as $m_1 \to \infty$.

The lower  branch, $E_- $, which starts from   $|m_1-m_2|$,  rises monotonically 
with increasing $\alpha$ to meet the upper branch at $E_c$. From $\ds ~ E_- = |m_1 - m_2| + {1\over 2} 
\left( {{m_1 m_2} \over{|m_1-m_2|}}\right) {\alpha^2 \over n^2}  + O(\alpha^4)~$ 
for $m_1 \ne m_2$, and $\ds ~ E_- = m \frac{\alpha}{n} + O(\alpha^3)~$ for $m_1=m_2=m$, 
 it is clear that the lower branch does not have the Balmer limit at small $\alpha$, 
and so is `unphysical' in 
this sense. The `unphysical' lower branch arises because  an `empty' vacuum was used, 
so that positive and negative energy solutions (rather than particles and antiparticles) are 
retained, and this means that solutions with two-body energies of the type  $m_1+m_2, m_1-m_2, 
-m_1+m_2, -m_1-m_2$ occur. However, the use of the empty vacuum is the price that 
needs to be  paid for obtaining analytic solutions.

 Interestingly, the same spectrum \re{1.20} was obtained previously by 
Berezin, Itzykson and Zinn-Justin
\cite{BIZ} as  poles of the scattering matrix in eikonal approximation, by Todorov \cite{Tod}, 
using a ``quasipotential"
 one-time reduction of the ladder Bethe-Salpeter equation, by Savrin and Troole \cite{ST}
as a solution of the ladder Bethe-Salpeter equation with retarding propagators, and 
 recently by Tretyak and Shpytko 
\cite{Sh} in a Fokker action formulation.  Berezin, Itzykson and Zinn-Justin state that their
analysis implies that \re{1.20} contains, 
in perturbation-theory language,  all recoil, ladder and crossed ladder effects, 
but no radiative corrections. However, as has been pointed out recently \cite{DD00},  the 
results analogous to \re{1.20} but for the massive chion  exchange case 
(specifically for $\mu / m = 0.15$), lie somewhat above the numerical Feynman-Schwinger 
calculations of Nieuwenhuis 
and Tjon \cite{NT}, which contain all effects save for radiative corrections. 

The method of  elimination of mediating fields, used in \cite{Dar98,BD98}, 
 is to some extend similar to that used in 
the formalism of Fokker action integrals in relativistic mechanics.  Although the Fokker action
formalism deals with a finite number of degrees of freedom,  in the case 
of scalar interactions its quantized counterpart \cite{Sh} 
leads to the same expression \re{1.20} for the energy as a function of quantum number $n$  
as that obtained in \cite{Dar98,BD98}. The difference
is in  different definitions of $n$ in the two treatments. This is due to different definitions 
of scalar interactions in the mechanical and QFT pictures, and by quantization ambiguities of 
the classical Fokker action.

 It  is evidently of interest to extend the method employed in \cite{Dar98,BD98} 
to include spinor particle fields.
As a first step 
 we consider a generalized Yukawa model consisting of fermions, described by a spinor field $\Psi$,
and of bosons, described by a scalar field $\varphi$, interacting via a 
 mediating scalar field $\chi$.
Our starting point is the Lagrangian density ($c=\hbar =1$)
\begin{equation}\label{1.1}
{\cal L}=\bar \Psi (i\gamma ^\nu\partial _\nu-g_1\chi -m_1)\Psi  + \partial _\nu\varphi ^*\partial ^\nu\varphi  -
m_2^2  \varphi ^* \varphi -g_2 \varphi ^* \varphi \chi 
+\frac12 \partial _\nu\chi\partial ^\nu\chi -\frac12\mu^2\chi^2
 \end{equation}

The fields of the model \re{1.1} satisfy the equations 
\begin{eqnarray}\label{1.2}
(i\gamma ^\nu\partial _\nu-g_1\chi -m_1)\Psi &=0,\\
\label{1.3}
(\partial _\nu\partial ^\nu+m_2^2+g_2\chi) \varphi  &=0,
\end{eqnarray}
and the conjugates of \re{1.2} and \re{1.3}, as well as
\begin{equation}\label{1.4}
(\partial _\nu\partial ^\nu+\mu^2)\chi=\rho, 
\end{equation}
with  
 \begin{equation}\label{1.5}
\rho=  -(g_1\bar \Psi \Psi +g_2\varphi ^*\varphi )
\end{equation}
Equation \re{1.4} has the formal solution 
\begin{equation}\label{1.6}
\chi =\chi_0+\chi_1,
\end{equation}
where
\begin{equation}\label{1.6-1}
\chi_1=\bra D*\rho\ket:=\int dx'D(x-x')\rho(x'),
\end{equation}
 $dx = d^Nx \,dt$ (in $N+1$ dimensions), and  $\chi_0 (x)$ satisfies the homogeneous 
(or free field) equation,  
\begin{equation}\label{1.7}
(\partial _\nu\partial ^\nu+\mu^2)\chi_0=0,
\end{equation}
 while $D(x-\xp)$ 
is a covariant Green function (or chion  propagator, in the 
 language of QFT), such that
\begin{equation}\label{1.8}
(\partial _\nu\partial ^\nu+\mu^2)D(x-x')=\delta^{N+1}(x-x').
\end{equation}

Equation \re{1.8} does not specify $D(x-\xp)$ uniquely since, for 
example, any solution of the homogeneous equation can be added 
to it without invalidating \re{1.8}. This allows for a certain freedom in the choice 
of $D (x)$,  as is discussed in standard texts (e.g. refs. \cite{Jack,Bar}).
Substitution of the formal solution \re{1.6} into Eqs. \re{1.2} and 
\re{1.3} yields the 
``reduced'' equations
\begin{eqnarray}\label{1.9}
\left(i\gamma ^\nu\partial _\nu-g_1(\chi_0+\chi_1) -m_1
\right)\Psi =0 ,\\
\label{1.10}
(\partial _\nu\partial ^\nu+m_2^2)\varphi=-g_2(\chi_0+\chi_1)\varphi .
\end{eqnarray}
These equations are  derivable from the action principle $\ds \delta 
\int dx\, {\cal L} = 0$, corresponding to the Lagrangian density
\begin{equation}
\label{1.13}
{\cal L}=\bar \Psi (i\gamma ^\nu\partial _\nu -m_1)\Psi 
 + \partial _\nu\varphi ^*\partial ^\nu\varphi  -
m_2^2 \varphi ^* \varphi +\chi_0\rho
+\frac12 \rho\bra D*\rho\ket
\end{equation}
provided that $D(x-\xp) = D(\xp-x)$. 

The QFTs based on \re{1.1} and \re{1.13} are equivalent in the sense that 
they lead to identical invariant 
matrix elements in various order of covariant perturbation theory.  
The difference is that, in the formulation based on \re{1.13}, 
the interaction 
term that contains the chion propagator $D(x-\xp)$  leads to
Feynman diagrams that correspond to 
processes involving virtual chions only. On the other hand,  
the interaction term that contains 
$\chi_0$ corresponds to Feynman 
diagrams that cannot be generated by the previous term,
such as those  with external (physical) chion lines. 

The reformulated Lagrangian  \re{1.13} contains two types of interactions: 
``local'' 
interactions, $\chi_0 \rho$, of the particle densities $\varphi^* (x) \varphi (x)$ 
and $\bar \Psi (x) \Psi (x)$
 with the free mediating field $\chi_0(x)$, and the ``nonlocal'' 
interaction, $\frac12 \rho\bra D*\rho\ket$, in which the 
chion propagator appears explicitly.  This may seem like a 
complication rather
 than simplification of the theory based on \re{1.1}. However, 
as we will show, the form
  \re{1.13} leads to a model for which exact eigenstates of the 
 Hamiltonian can be obtained.


\section{Feshbach-Villars Formulation of the Scalar Field}

\setcounter{equation}{0}
\renewcommand{\theequation}{2.\arabic{equation}}

We rewrite the scalar field $\varphi$ of this model in the Feshbach-Villars (FV) 
formulation \cite{FV}.  The reason for doing so is that this leads to 
a QFT Hamiltonian which is Schr\"odinger-like in form, 
for which exact eigensolutions can be obtained. 
In the FV formulation,  the  field $\varphi$ and its time-derivative 
$\dot \varphi$ are replaced by a two-component vector which is defined as
\begin{equation}\label{2.1}
\phi=\left[\matrix{\phi_1 = {1\over {\sqrt{2m_2}}}(m_2 \varphi + i \dot \varphi)&\cr
\phi_2 = {1\over {\sqrt{2m_2}}}( m_2 \varphi - i \dot \varphi)&\cr}\right]  \;\;\; 
\end{equation}
so that, for example, $2 m_2 \,\varphi^* \varphi = (\phi_1^* + \phi_2^*)(\phi_1+\phi_2) 
= \phi^{\dag} \eta \tau \phi$, where $\eta$ and $\tau$ are the matrices
\begin{equation}\label{2.2}
\eta = \left [\matrix{1&0\cr 0& -1\cr}\right ]\;\;\;\; {\rm and} 
\;\;\;\;\tau = \left [\matrix{1&1\cr -1& -1\cr}\right ] .
\end{equation}
In the FV formulation the equation of motion \re{1.3} takes on the form
\begin{equation}\label{2.3}
i \dot \phi = - {1\over {2m_2}} \nabla^2 \tau \phi + m_2 \eta \phi + 
{g_2 \over {2m_2}} \tau \phi \chi, 
\end{equation}
or, upon using \re{1.6}, the form
\begin{equation}\label{2.4}
i \dot \phi = - {1\over {2m_2}} \nabla^2 \tau \phi + m_2 \eta \phi + 
{g_2\over {2m_2}} \tau \phi (\chi_0+ \bra D*\rho\ket),
\end{equation}
where $\ds \rho =  -(g_1\bar \Psi \Psi +{g_2\over{2m_2}} \phid  \eta \tau \phi)$.

 Equations \re{2.4} and   \re{1.9} are derivable from the 
Lagrangian density 
\begin{equation}
\label{2.5}
{\cal L}_{FV}=\bar \Psi (i\gamma ^\nu\partial _\nu -m_1)\Psi 
 + 
i \phi^{\dag}(x) \eta \dot \phi(x)  - 
{1\over {2m_2}} \nabla \phib(x) \cdot \nabla \phi (x) - m_2 \phi^{\dag} 
(x)\phi (x)  
 +\chi_0\rho
+\frac12 \rho\bra D*\rho\ket ,
\end{equation}
where $\phib = \phi^{\dag} \eta \tau \phi$.


\section{Quantization}

\setcounter{equation}{0}
\renewcommand{\theequation}{3.\arabic{equation}}

The momenta corresponding to $\phi_1$ and $\phi_2$ are 
\begin{equation}\label{3.1}
p_{\phi_1} = {\di {\cal L}_{FV} \over {\di \dot {\phi_1}}} = i \phi_1^*,~~~~
p_{\phi_2} = -i \phi_2^*
\end{equation}
Thus,  the Hamiltonian density is given by the expression
\begin{equation}\label{3.2}
{\cal H}(x) = \Psi^\dag(x)  {\hat h_1} (x) \Psi(x)+
 \phi^{\dag}(x) \eta {\hat h_2} (x) \phi(x) - \chi_0(x)\rho(x) -
\frac12 \rho(x)\bra D*\rho\ket
\end{equation}
where $\ds {\hat h_1}(x)=-i\vec \alpha  \cdot  \nabla +m_1\beta ,~
{\hat h_2}(x) = \tau \left(-{1\over {2m_2}} \right) \nabla^2 + m_2 \,\eta$, and where we
 have suppressed  terms like \hfill \break
 $\nabla \cdot (\phib(x) \nabla \phi(x))$ that
 vanish upon integration and application of Gauss' theorem.
We use canonical equal-time quantization,  whereupon the non-vanishing 
anticommutation relations, and commutation relations, are
\begin{equation}\label{3.3}
\{\Psi_\alpha(\bx,t) ,\Psi_\beta(\by,t) ^\dag \}=\delta _{\alpha \beta } \delta^N (\bx-\by),
~~~\alpha ,\beta =1...4;~~~
  [\phi_a(\bx,t),\phi^{\dag}_b(\by,t)] =  
 \eta_{ab}\delta^N(\bx-\by),
 \;\;\; a,b=1,2
\end{equation}
where $\eta_{ab}$ are 
elements of the $\eta$ matrix \re{2.2}.
Using these commutation relations, the Hamiltonian operator 
can be written as
\begin{equation}\label{3.3-1}
H = \int d^N x\, [{\cal H}_0(x) + {\cal H}_{\chi} (x) + {\cal H}_I 
(x) ], 
\end{equation}
where (suppressing the Hamiltonian of the free chion field),
\begin{eqnarray}\label{3.4}
{\cal H}_0(x) &=&\Psi^\dag(x)  {\hat h_1} (x) \Psi(x)+
 \phi^{\dag}(x) \eta {\hat h_2} (x) \phi(x),\\
{\cal H}_\chi (x)&=&-\chi_0(x)\rho(x) 
\end{eqnarray}
and 
\begin{eqnarray}\label{3.5}
;{\cal H}_I (x); \;\;\; =
 &-&{g_1^2 \over {2}} \int d\xp \,D(x-\xp)\, \Psib (x)(\Psib (\xp) 
\Psi(\xp))\Psi(x)\nn\\
&-& {g_2^2 \over {8m_2^2}} \int d\xp \,D(x-\xp)\, \phib (x)(\phib (\xp) 
\phi(\xp))\phi(x) \nn\\
&-& {{g_1g_2} \over {4m_2}} \int d\xp \,D(x-\xp)\, \phib (x)(\Psib (\xp) 
\Psi(\xp))\phi(x) \nn\\
 &-&  {{g_1g_2} \over {4m_2}}\int d\xp \,D(x-\xp)\, \Psib (x)(\phib (\xp) 
\phi(\xp))\Psi(x),
\end{eqnarray}
and where we have used the commutation relations 
\re{3.3}  to reorder $\phib (x) 
\phi(x) \phib (\xp) \phi (\xp)$ as $ \phib(x)(\phib(\xp) 
\phi(\xp))\phi(x)$ and    $\Psib (x) 
\Psi(x) \Psib (\xp) \Psi (\xp)$ as $ \Psib(x)(\Psib(\xp) 
\Psi(\xp))\Psi(x)$. Note 
that no infinities are dropped upon performing the `normal ordering' of scalar field operators, 
since none 
arise on account of the property that $\tau^2 = 0$.  However, in the case of the spinor field, 
 reordering yields an infinite constant which can be absorbed into  the  total energy of the system.  
Of course, one can simply start from the normal-ordered Hamiltonian.
 We stress that this `normal ordering' 
is not the same as the conventional one \cite{BjD}, since we normal order the entire field operators, 
$\phi, \psi$ and $\phi^{\dag}, \psi^{\dag}$, not  positive and negative frequency parts individually. 
For this reason, we denote it as $;H;$ rather than as $:H:$.

As already mentioned, ${\cal H}_I$ contains the covariant chion 
propagator, hence in conventional covariant perturbation theory it 
leads to Feynman diagrams with internal chion lines. On the other 
hand, ${\cal H}_{\chi}$ leads to Feynman diagrams with external 
chions.  However, we shall not pursue covariant perturbation theory 
in this work, and so shall not consider that approach further. Rather, 
we shall consider an approach that leads to some exact eigenstates of 
the Hamiltonian \re{3.3-1}, but with ${\cal H}_\chi = 0$.


\section{ Truncated model}

\setcounter{equation}{0}
\renewcommand{\theequation}{4.\arabic{equation}}

In what follows we shall consider the  truncated model for which the term
 ${\cal H}_{\chi}$ in \re{3.3-1} is suppressed. Such a 
Hamiltonian is appropriate for describing systems for which there is 
no annihilation or decay into chions, or chion-phion/psion scattering.

In the Schr\"odinger picture we can take $t=0$. Therefore, we shall 
use the notation that, say  $\phi(\bx,t=0) = \phi(\bx)$, etc., for the 
QFT operators.
This allows us to express the interaction part of the Hamiltonian \re{3.5} as
\begin{eqnarray}\label{4.1}
;{\cal H}_I (x); \;\;\; =
 &-&{g_1^2 \over {2}} \int d^N\xp \,G(\bx-\bxp)\, \Psib (\bx)(\Psib (\bxp) 
\Psi(\bxp))\Psi(\bx)\nn\\
&-& {g_2^2 \over {8m_2^2}} \int d^N\xp \,G(\bx-\bxp)\, \phib (\bx)(\phib (\bxp) 
\phi(\bxp))\phi(\bx) \nn\\
&-& {{g_1g_2} \over {4m_2}} \int d^N\xp \,G(\bx-\bxp)\, \phib (\bx)(\Psib (\bxp) 
\Psi(\bxp))\phi(\bx) \nn\\
&-&  {{g_1g_2} \over {4m_2}}\int d^N\xp \,G(\bx-\bxp)\, \Psib (\bx)(\phib (\bxp) 
\phi(\bxp))\Psi(\bx),
\end{eqnarray}
where
\begin{equation}\label{4.2}
G(\bx-\bxp) = \int_{-\infty}^{\infty} D(x-\xp)\, dt^{\prime}= 
{1\over {(2\pi)^N}} \int d^Np\, e^{i\bp \cdot (\bx-\bxp)} {1\over 
{\bp^2 + \mu^2}}\,.
\end{equation}
Explicitly, for $N=3$ spatial dimensions this becomes
\begin{equation}\label{4.3}
G(\bx-\bxp) = {1\over {4\pi}} {e^{-\mu |\bx-\bxp|} \over {|\bx-\bxp|} }, 
\end{equation}
for $N=2$ it is 
\begin{equation}\label{4.4}
G(\bx-\bxp) = {1\over {2\pi}} k_0(\mu |\bx-\bxp|),
\end{equation}
where $k_0(z)$ is the modified Bessel function,
whereas for $N=1$ it has the form
\begin{equation}\label{4.5}
G(x-\xp) = {1\over {2\mu}} e^{-\mu|x-\xp|}
\end{equation}


\section{Empty vacuum and one-particle eigenstates}

\setcounter{equation}{0}
\renewcommand{\theequation}{5.\arabic{equation}}

As in Refs. \cite{Dar98,BD98}, 
we define an empty vacuum state, $|\tilde 0\rangle$, such that 
\begin{equation}\label{5.1}
\phi_a |\tilde 0\rangle =  \Psi_\alpha  |\tilde 0\rangle = 0. 
\end{equation}
This is different from the ``Dirac vacuum" $|0\rangle$ 
of conventional QFT, which is annihilated 
by only the positive frequency part of $\varphi$ and $\Psi$ and by the negative 
frequency parts of $\varphi^{\dag}$ and $\Psi^{\dag}$. 

With the definition \re{5.1}, the one-particle scalar state defined as
\begin{equation}\label{5.2}
|1_{\phi} \rangle = \int d^Nx\,\phid (\bx) \eta f(\bx) |\tilde 0
\rangle, 
\end{equation}
where $f(\bx)$ is a two-component vector,
is an eigenstate of the truncated QFT Hamiltonian  (${\cal H}_\chi 
= 0$) with eigenvalue $E_1$ provided that the $f(\bx)$ is a solution 
of the equation
\begin{equation}\label{5.3}
{\hat h_2}(\bx) f(\bx) = E_1 f(\bx). 
\end{equation}
This is just the free-particle Klein-Gordon equation  for stationary 
states (in Feshbach-Villars form). It has, of course, 
all the usual negative-energy  ``pathologies'' of the KG equation. 
The presence of negative-energy solutions is a consequence of the 
use of the empty vacuum.

Similarly,  the state 
\begin{equation}\label{5.4}
|1_\Psi \ket=\int d^3x\,\tilde F_{\alpha}(\bx)\,\Psi^{\dag}_{\alpha}(x)                     
|{\widetilde 0}\ket ,
\end{equation}
is an eigenstate of $;H;$ with eigenenergy $E_1$, provided               
that the four coefficient amplitudes $\tilde F_{\alpha}(\bx)$ are                  
solutions of                                                                    
\begin{equation}\label{5.5}
\{[h_1(\bx)]_{\alpha\,\beta}-                                               
{E}_1\,\delta_{\alpha\,\beta}\}\tilde F_{\beta}(\bx)=0, 
\end{equation}
or                                                                              
\begin{equation}\label{5.6}
(h_1(\bx)-E_{1})\tilde F(\bx)=0
\end{equation}
in matrix notation, that is provided the spinor $\tilde F$ is a solution               
of the usual Dirac eigenvalue equation \re{5.6}. 
Note that summation on repeated spinor indices $\alpha, \beta$ is 
implied in equations \re{5.4} and \re{5.5}.


\section{Two-particle eigenstates}

We consider a mixed two-particle phion-plus-psion system described by 
\setcounter{equation}{0}
\renewcommand{\theequation}{6.\arabic{equation}}
\begin{equation}\label{6.1}
|2\rangle = \int d^Nx\,d^Ny\; F_{\alpha \, j}(\bx,\by)\;\Psid_ \alpha (\bx) 
\phid_j(\by) |\tilde 0\rangle
\end{equation}
where summation on  repeated spinor and boson indices, $\alpha $ and $j$, is implied. This state 
is an exact eigenstate of the truncated QFT Hamiltonian (\re{3.3-1} with  ${\cal H}_
\chi = 0$) provided that the $4\times 2$  coefficient matrix $F = 
[F_{\alpha \,j}]$ is a solution of the two-body equation,
\begin{equation}\label{6.2}
 {\hat h_1}(\bx)  F  + \left[\eta {\hat h_2}(\by) \eta 
F^{\rm T }(\bx,\by)\right]^{\rm T } + V(\bx-\by) \gamma ^0\left[\eta\tau\eta
 F^{\rm T}(\bx,\by)\right]^{\rm T } = EF
\end{equation}
where the superscript T stands for ``transpose''. The potential 
here is given by
\begin{equation}\label{6.3}
V(\bx-\by) = -\frac{ g_1 g_2}{2 m_2} G(\bx-\by)
\end{equation}
where $G(\bx - \by)$ is specified in equations \re{4.2}-\re{4.5}.
Alternatively, we can regard $|2\ket$ as a variational approximation to the eigenstate of the 
complete Hamiltonian \re{3.3-1}, since $\bra 2 | H_\chi | 2 \ket =0 $.

For $|2\ket$ to be an eigenstate of the momentum operator with                     
eigenvalue $\bP_{TOTAL}=0$, it is necessary that the 4$\times$2 `hyperspinor'
$F(\bx,\by)$ be of the form $F(\br)$,    where $\br=\bx-\by$.                
Let us define 
\begin{equation}\label{6.5}
F(\br)=\left [ \matrix {s (\br)\cr t  (\br) }\right ],
\end{equation}
where, in $N=3$ spatial dimensions, 
\begin{equation}\label{6.4}
s (\br) = \left [ \matrix {
F_{11}(\br) & F_{12}(\br)\cr F_{21}(\br) & F_{22}(\br)\cr
} \right ],
~~~~ t (\br) = \left [ \matrix {F_{31}(\br) & F_{32}(\br)\cr F_{41}(\br) & F_{42}(\br)\cr} 
\right ].
\end{equation}
Then equation \re{6.2} can be written as two coupled equations 
 for the 2$\times$2-matrices $s(\br),~t(\br)$
\begin{eqnarray}\label{6.6}
-i\bsi \cdot \bnab t(\br) +
 \left[\eta\Big(\hat h_2(\br)+ \tau V(\br) \Big)\eta \, s ^{\rm T}(\br) \right]^{\rm T}
+(m_1-E) s(\br) =0,\\
\label{6.7}
-i\bsi \cdot \bnab s(\br) +
 \left[\eta\Big(\hat h_2(\br) - \tau V(\br) \Big) \eta \,  t ^{\rm T} (\br) \right]^{\rm T}
-(m_1+E) t(\br)=0.
\end{eqnarray}
In the absence of interactions ({\sl i.e.} $ V = 0$), these equations have solutions with the 
following four types of energy eigenvalues: $\omega(\bp,m_1) + \omega(\bp,m_2)$,  
$-\omega(\bp,m_1) + \omega(\bp,m_2)$, $\omega(\bp,m_1) - \omega(\bp,m_2)$,
$-\omega(\bp,m_1) - \omega(\bp,m_2)$, where $\omega(\bp,m) = \sqrt{\bp^2+m^2}$.
 The first and last are positive  and negative energy 
eigenvalues, respectively, while the middle two are `mixed'. The appearance of the negative energies 
is a reflection of our use of the empty vacuum state, as discussed earlier.


\section{$J^P$ eigenstates and radial reduction in $N=3$ dimensions}

\setcounter{equation}{0}
\renewcommand{\theequation}{7.\arabic{equation}}

The Hamiltonian \re{3.3-1} of the theory commutes with the total angular momentum and 
parity operators.
If the two-particle state \re{6.1} is to be an eigenstate of $J_3$  where
\begin{equation}\label{7.1}
\bJ= \int
d^3x\,\Psi^{{\dag}}(\bx,t)\,{\bj}(\bx)\,\Psi(\bx,t) 
+\int
d^3x\,\phi^{\dag}(\bx,t)\eta\,\bl(\bx)\,\phi(\bx,t),
\end{equation}
with $\bj(\bx)=\bl(\bx)+\bs=-i\bx\times\bnab_{\bx}+\frac12\bsi$, then we require that 
the hyperspinor $F$ (cf. Equation \re{6.1}) must satisfy the equation
\begin{equation}\label{7.2}
(j_3(\bx)+l_3(\by))F(\bx,\by)=m_JF(\bx,\by)
\end{equation}
In other words, in the rest frame where $\bP_{TOTAL}|2\ket= 0\;|2 \ket$, we require that
\begin{equation}\label{7.3}
\left( l_3(\br)+{1\over 2}\sigma _3 \right)F(\br)=m_jF(\br).
\end{equation}
 In a similar fashion $\bJ^2|2 \ket= j(j+1)|2 \ket$ implies that the
matrix $F$ in the rest frame must satisfy the equation
\begin{equation}\label{7.4}
\left( \bl^2+{3\over 4} +\bl\cdot\bsi \right) F(\br)=j(j+1)F(\br)
\end{equation}
The  components $s,~t$ of $F$ satisfy Eqs. \re{7.3}, \re{7.4} individually.
 
For $|2\ket$ to be a parity eigenstate, and since $V(r)$ is invariant 
under space reflection
the matrix $F(\br)$ must have the property that
\begin{equation}\label{7.5}
\beta F(\br)=\pm F(-\br), 
\end{equation}
where the $\pm$ are the parity quantum numbers.
This means that
\begin{equation}\label{7.6}
s(-\br)=\pm s(\br),~~~~t(-\br) =\mp t(\br).
\end{equation}
From Eqs. \re{7.3} - \re{7.6} it follows that
\begin{equation}\label{7.7}
F (\br)= 
\left [ \matrix {s (\br)\cr t (\br) }\right ]=
\left [ \matrix {
k_1(r)\zeta_{j,m_j}^{\ell=j\pm\frac12}(\hat{\br}) & 
-k_2(r)\zeta_{j,m_j}^{\ell=j\pm\frac12}(\hat{\br})\cr
q_1(r)\zeta_{j,m_j}^{\lambda=j\mp\frac12}(\hat{\br}) &
-q_2(r)\zeta_{j,m_j}^{\lambda=j\mp\frac12}(\hat{\br})\cr
} \right ],    ~~~~ \hat{\br} = \frac{\br}{r},
\end{equation}
where the upper sign corresponds to parity `+', and the lower to parity `-',
and $k_1,~k_2,~q_1,~q_2$    are radial functions.
The normalized ``spinor harmonics" $\zeta_{j,m_j}^{\ell=j\pm\frac12}(\hat {\br})$ 
are, explicitly, 
\begin{equation}\label{7.8-1}
\zeta_{j,m_j}^{\ell=j-\frac12}(\hat {\br})=\frac{1}{\sqrt{2\ell+1}}
\left [ \matrix {
\sqrt{\ell+m_j+\12}Y_{\ell}^{m_j-\frac12}(\hat {\br})\cr
\sqrt{\ell-m_j+\12}Y_{\ell}^{m_j+\frac12}(\hat {\br})\cr
} \right ],
\end{equation}
and
\begin{equation}\label{7.8-2}
\zeta_{j,m_j}^{\ell=j+\frac12}(\hat{\br})=\frac{1}{\sqrt{2\ell+1}}
\left [ \matrix {
\sqrt{\ell-m_j+\12}Y_{\ell}^{m_j-\frac12}(\hat{\br})\cr
-\sqrt{\ell+m_j+\12}Y_{\ell}^{m_j+\frac12}(\hat{\br})\cr
} \right ],
\end{equation}

Finally, substituting \re{7.7} into \re{6.6}, \re{6.7} we find that the radial functions  must satisfy
 the following system of four  equations:
\begin{eqnarray}
\label{7.9}
&&\hat \Pi _{\sigma}^{ \mp}q_1+\frac{\hat \Pi _{\pm}^2}{2m_2}(k_1+k_2)+(m_1+m_2)k_1 + V(r)(k_1+k_2)=Ek_1,\nn\\
&&\hat \Pi _{\sigma}^{ \mp}q_2-\frac{\hat \Pi _{\pm}^2}{2m_2}(k_1+k_2)+(m_1-m_2)k_2 - V(r)(k_1+k_2)=Ek_2,\nn\\
&&\hat \Pi _{\sigma}^{\pm}k_1+\frac{\hat \Pi _{\mp}^2}{2m_2}(q_1+q_2)+(-m_1+m_2)q_1 - V(r)(q_1+q_2)=Eq_1,\\
&&\hat \Pi _{\sigma}^{\pm}k_2-\frac{\hat \Pi _{\mp}^2}{2m_2}(q_1+q_2)-(m_1+m_2)q_2 + V(r)(q_1+q_2)=Eq_2,\nn
\end{eqnarray}
where $\ds V(r) = - \alpha \frac{e^{-\mu r}}{r}$ with $\ds \alpha = \frac{g_1 g_2}{8 \pi m_2}$,
 and where we have introduced the operators
\begin{eqnarray}\label{7.10}
&&\hat \Pi _{\sigma}^{ \pm}:=-\frac{i}{r}\left(r\frac{d}{dr}+[1\pm(j+\12)]\right)\nn\\
\\
&&\hat \Pi _{\pm}^2:= -\frac1r\left(r\frac{d^2}{dr^2}+2\frac{d}{dr}-\frac{(j+1/2)(j+1/2\pm1)}{r}\right). \nn
\end{eqnarray}
If we let $E=\varepsilon + m_1$ and consider the $m_1 \to \infty$ limit of \re{7.9}, we 
find that $q_1, q_2 \to 0$ while $k_1+k_2$ satisfies the radial 
Klein-Gordon equation (with scalar coupling).
 Similarly, if $m_1$ is replaced by $m_2$ in this procedure, then $k_2, q_2 \to 0$ while 
$q_1$ and $- i k_1$ satisfy the radial Dirac equations (with scalar coupling). Thus, equations 
\re{7.9} have the expected one-body limits.

Defining $q_\pm=q_1\pm q_2,$ and $k_\pm=k_1\pm k_2$, then adding the first two, and 
separately the last two, of equations \re{7.9}, we obtain
\begin{eqnarray}\label{7.11}
-i\frac{dq_+}{dr}=(E-m_1)k_+-m_2k_-+\frac{i}{r}[1\mp(j+\12)]q_+\nn\\
\\	
-i\frac{dk_+}{dr}=(E+m_1)q_+-m_2q_-+\frac{i}{r}[1\pm(j+\12)]k_+\nn
\end{eqnarray}
Using definitions \re{7.10}, differentiating \re{7.11}, and substituting the result into \re{7.9} we obtain
\begin{eqnarray}\label{7.12}
{-}\frac{i}{2m_2}\left[ (E+m_1+m_2)q_+'+ \frac{m_2[1\mp(j+\12)]}{r}(q_++q_-)\right]-
\frac{1{\mp}(j{+}\12)}{2m_2r}k_+'+\qquad\qquad\nn\\
\left[  \frac{(j{+}\12)^2-1}{2r^2 m_2}+V\right] {k_+} +
\frac{m_1{+}m_2{-}E}{2}(k_+{+}k_-)=0\\
\label{7.13}         
{-}\frac{i}{2m_2}\left[ (m_2-m_1-E)q_+'+ \frac{m_2[1\mp(j+\12)]}{r}(q_+-q_-)\right]+
\frac{1{\mp}(j{+}\12)}{2m_2r}k_+'-\qquad\qquad\nn\\
\left[  \frac{(j{+}\12)^2-1}{2r^2 m_2}+ V\right] {k_+} +
\frac{m_1{-}m_2{-}E}{2}(k_+{-}k_-)=0\\
\label{7.14}         
{-}\frac{i}{2m_2}\left[ (m_2-m_1+E)k_+'+ \frac{m_2[1\pm(j+\12)]}{r}(k_++k_-)\right]-
\frac{1{\pm}(j{+}\12)}{2m_2r}q_+'+\qquad\qquad\nn\\
\left[  \frac{(j{+}\12)^2-1}{2r^2 m_2}-V\right] {q_+} +
\frac{-m_1{+}m_2{-}E}{2}(q_+{+}q_-)=0\\
\label{7.15}         
{-}\frac{i}{2m_2}\left[ (m_2+m_1-E)k_+'+ \frac{m_2[1\pm(j+\12)]}{r}(k_+-k_-)\right]+
\frac{1{\pm}(j{+}\12)}{2m_2r}q_+'-\qquad\qquad\nn\\
\left[  \frac{(j{+}\12)^2-1}{2r^2 m_2}-V\right] {q_+} -
\frac{m_1{+}m_2{+}E}{2}(q_+{-}q_-)=0,
\end{eqnarray}
where $\ds q_+' = \frac{dq_+}{dr}$, {\sl etc.}
Eqs. \re{7.12} - \re{7.15} do not contain $q_-',~k_-'$. Therefore, solving \re{7.11} for 
$q_-,~k_-$, putting $Q_+=-ik_+$,
 and substituting into \re{7.12}, \re{7.14} (or  into  \re{7.13}, \re{7.15}) 
leads to a system of two radial Dirac-like equations: 
\begin{eqnarray}\label{7.16}
Q_+'+\frac{1\pm(j+\12)}{r}Q_+-\frac{m_2}{E} V(r) q_+-\epsilon _2 q_+=0, \nn \\
\\
-q_+'-\frac{1\mp(j+\12)}{r}q_++\frac{m_2}{E} V(r) Q_+-\epsilon _1 Q_+=0, \nn
\end{eqnarray}
where
\begin{equation}\label{7.17}
\epsilon _1= \frac{(E-m_1-m_2)(E-m_1+m_2)}{2E},~
\epsilon _2=\frac{(E+m_1+m_2)(E+m_1-m_2)}{2E}
\end{equation}
If we make the replacement, $E=m_1+m_2+\varepsilon$, and assume that 
$|\varepsilon|,|V|<<m_1,m_2$,  then equations  \re{7.16} reduce to 
\begin{eqnarray}\label{7.18}
Q_+''-\frac{\kappa(\kappa+1)}{r^2}Q_+
+2m_r(\varepsilon- V(r))Q_+=0,~~~~m_r= 
\frac{m_1m_2}{m_1+m_2}, 
\end{eqnarray}
which is just the reduced radial Schr\"odinger equation for the relative motion of the
 two particles of masses $m_1$ and $m_2$, with $\ell=j+1/2$ if $\kappa =j+1/2=\ell$ and 
$\ell=j-1/2$ if $\kappa =-(j+1/2)=-(\ell+1)$. Thus, the positive-energy solutions  have
 the expected non-relativistic limit. On the other hand, for the negative-energy solutions, if  
$E=-(m_1+m_2+\varepsilon)$, and $|\varepsilon|,|V|<<m_1,m_2$, equations  \re{7.16} reduce to
\begin{eqnarray}\label{7.19}
q_+''-\frac{\kappa(\kappa-1)}{r^2}q_+
+2m_r(\varepsilon+ V(r))q_+=0,
\end{eqnarray}
which is  the  radial Schr\"odinger equation, with $\ell=j+1/2$ if $\kappa =-(j+1/2)=-\ell$ and 
$\ell=j-1/2$ if $\kappa =j+1/2=\ell+1$. We note that the sign of $V(r)$ is effectively
 reversed for the negative-energy  solutions,
relative to that for the positive-energy solutions of equation \re{7.18}.  
This means that if there are positive-energy bound states, then there are no bound states 
for the negative-energy solutions or vice versa. This is  similar to that, which occurs 
for the scalar Yukawa model \cite{Dar98,BD98}. 

We also note that for the mixed-energy solutions of the type 
$E=m_1-m_2-\varepsilon$, where we take $m_1 > m_2$ for
 definiteness, we find that the corresponding non-relativistic 
reduction leads to an equation exactly like \re{7.18} but with 
an unphysical reduced mass $m_r = m_1 m_2/(m_1-m_2)$. 
This indicates that \re{7.16} also admit `unphysical' 
mixed-energy solutions, in addition to the positive and negative 
energy solutions. This is consistent with our earlier observation in VI  
regarding the solutions for the `free particle' case with V=0.

The coupled equations \re{7.16} can be solved readily, for arbitrary mass $\mu$ of the
 mediating scalar field, by standard numerical techniques.  Once $Q_+, q_+$ have been 
obtained, $k_-$ and $q_-$ can be determined from \re{7.11}. This, then, determines 
$F(\br)$ completely, for any $J^P$ state.


\section{Two-body bound states in 3+1 for massless chion exchange.}

\setcounter{equation}{0}
\renewcommand{\theequation}{8.\arabic{equation}}

We consider the solution of equations \re{7.16} for $N=3$ and massless chion exchange (i.e. $\mu =0$),
 in which case the 
interparticle potential is  Coulombic ({\i.e. $\ds V(r) = - \frac{g_1 g_2}{8\pi m_2} \frac{1}{r}$),
 and analytic solutions for the eigenvalues and 
eigenfunctions can be obtained. Thus, putting
\begin{eqnarray}\label{8.1}
Q_+=e^{-\beta r}\sum_{\nu=0}a_j r^{\gamma-1+\nu},~~~~~
q_+=e^{-\beta r}\sum_{\nu=0}b_j r^{\gamma-1+\nu} ,
\end{eqnarray}
we obtain
\begin{equation}\label{8.20}
(\kappa + \gamma + \nu) a_\nu + {\tilde \alpha} b_\nu - \beta a_{\nu-1} - \epsilon_2 b_{\nu-1} = 0
~~~~ 
(\kappa - \gamma - \nu) b_\nu - {\tilde \alpha} a_\nu + \beta b_{\nu-1} - \epsilon_1 a_{\nu-1} = 0,
\end{equation}
where 
\begin{equation}\label{8.21}
{\tilde \alpha} = \alpha {m_2 \over E}, ~~~~~~ \alpha =\frac{g_1g_2}{8\pi m_2},~~~~~~
\kappa =\pm \left(j+\12\right),~~~~ {\rm and} ~~~~ \nu = 0,1,2,...
\end{equation}
The case $\nu = 0$, with $a_{\nu-1}=b_{\nu-1}=0$, gives $\gamma = \sqrt{\kappa^2 +{\tilde 
\alpha}^2 }$.  

In addition, for $Q_+, q_+$ to be well behaved at infinity, the series \re{8.1} must terminate 
at $\nu = n' \ge 0$.  Then \re{8.20}, with $\nu = n'+1$ and $a_{n'+1}=b_{n'+1}=0$, 
gives the energy spectrum formula
\begin{equation}\label{8.2}
 n'+\gamma = \frac{m_1 {\tilde \alpha}}{\beta}
\qquad  n' = 0, 1, 2, ...\qquad {\rm where}
 \qquad \beta=\sqrt{-\epsilon _1\epsilon _2 }~ .
\end{equation}            
Note that the positive square roots  must be chosen for $\gamma$ 
and $\beta$ in order that the wave 
functions be well behaved at the origin and at infinity.   Since $n' \ge 0$ and $ \gamma, \beta$
are positive, it follows from \re{8.2} that $\tilde \alpha$ must be positive, which, from its 
definition in \re{8.21}, means that for $g_1 g_2 >0$ ({\sl i.e.} an attractive potential 
$V(r) = - \alpha / r$ ), the energy eigenvalue $E$ must be positive. This, together 
with  the requirement that $-\epsilon_1 \epsilon _2 > 0$ implies that the bound state energy spectrum
for this fermion-scalar system with {\sl scalar} Coulombic coupling must lie in the 
domain $ |m_1-m_2| < E < m_1+m_2$, exactly as for the scalar Yukawa model \re{1.20} 
\cite{BD98}. 

It is convenient to rewrite \re{8.2} in the form
\begin{equation}\label{8.3}                        
n'+\sqrt{\kappa ^2+\frac{\alpha^2m_2^2 }{E^2}}=\frac{\alpha}{\sqrt{1-w^2}},
~~~{\rm where}~~~w=\frac{E^2-m_1^2-m_2^2}{2m_1m_2}.
\end{equation} 
The eigenvalue spectrum equation \re{8.3} is not symmetric with respect to the particle masses. 
This reflects 
the different nature of the particles. The mass $m_1$ belongs to the spinor particle 
and the mass $m_2$ corresponds to the scalar one.			
Note that by  putting $\tilde n =n'+\sqrt{\left(j+\12\right)^2+\frac{\alpha^2m_2^2 }{E^2}}$ 
one can rewrite \re{8.3} in the form 
\begin{equation}\label{8.5}
E^2=m_1^2+m_2^2 \pm  2 m_1 m_2 \sqrt{1-\frac{\alpha ^2}{\tilde n^2}}
\end{equation}		
which formally coincides with the energy spectrum \re{1.20} for two scalar particles with 
scalar interaction obtained in Refs.  
\cite{Dar98,BD98}.
But here the `quantum number' $\tilde n$  depends on the energy.


Equation \re{8.3} is  exact  and  gives  the  energy spectrum of the two particle (scalar plus 
fermion) system. 
However the explicit determination of the function  $E=E(\alpha,m_1,m_2 )$ for 
various states requires the  solution of an
algebraic equation of  the sixth order in $E^2$. By contrast, the inverse  
dependence  $\alpha =\alpha(E,m_1,m_2)$ is relatively simple:
\begin{equation}\label{9.1}
\alpha = \frac{2m_1E\sqrt{1-w^2}\left( 2m_1n'E + \sqrt{\kappa^2(E^2+m_1^2-m_2^2)^2+4m_1^2
m_2^2n'^2(1-w^2)}
\right)}
{(E^2+m_1^2-m_2^2)^2}
\end{equation}
This solution, though analytic, is not particularly transparent, except for some special cases, such as the
equal-mass states with $n'=0$ ({\sl i.e.} $n=n'+ j+1/2 = j+1/2$), for which
\begin{equation}\label{9.30}
\alpha = 2 n \sqrt{1- \left(\frac{E}{2m}\right)^2 } \qquad {\rm or} \qquad 
E = 2 m \sqrt{1 - \left( \frac{\alpha}{2n} \right)^2 } ~~~~ n=j + \frac{1}{2} ,
\end{equation}
where $m=m_1=m_2$. 
This shows that equal-mass bound states are possible only for $\alpha < 2 n$ for 
 $n'=0$ ({\sl i.e.} $n = j+\frac{1}{2}$)
states, and that $E=0$ at the critical value of $\alpha_c=2 n$. However, for $n' >0$ states of the 
equal-mass case, the shape of $E(\alpha)$ is quite unlike the quarter circle \re{9.30}. Rather, 
$E(\alpha)$ decreases monotonically from $E(0)= 2 m$ towards zero, as $\alpha \to \infty$.

As an example, in Figure 1 we plot $\ds \frac{\alpha(E)}{n}$ 
for $n'=0$ states (for which $n= j + 1/2$), for three different
 mass combinations.  The equal-mass  curve is labelled  
$m_1/m_2=1$, and corresponds to the quarter circle of \re{9.30}.  The  deformed semi-circle corresponds to 
 $m_1/m_2 = 2$.  The apex of this curve is the critical 
value of $\alpha$ beyond which there are no real solutions 
for the two-particle bound state mass E. The two-branch curve 
straddling the vertical asymptote corresponds to 
 $m_1/m_2 = 1/2$. The vertical asymptote occurs 
at $E/m_2 = \sqrt{1- (m_1/m_2)^2}= \sqrt{3}/2$. There is a real solution
$E$  for any value of $\alpha$ for this $m_1/m_2 < 1$ case. 
Notice that every one of the curves lies in the domain 
$|m_1-m_2| < E < m_1 + m_2$ .


\epsfxsize=7.5cm
\begin{figure}[h]
\leavevmode
\begin{center}
{\epsffile{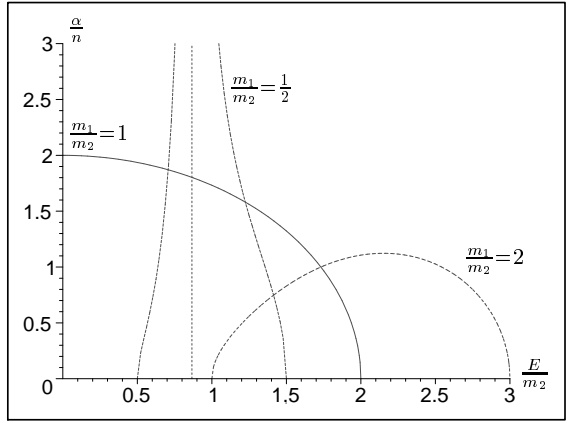}}
\end{center}
\caption{ Plot of $\alpha / n$ versus $E / m_2$ for $n' = 0~ (n = j + 1/2)$ states
of three different mass combinations, where $m_1$ is the fermion mass and
$m_2$ is the boson mass.     }
\end{figure}

We can, of course, evaluate and plot $\alpha(E)$ for any $n', j$ state and any values of $m_1, m_2$, 
using the analytic formula \re{9.1}. However, it is instructive to outline the general behaviour 
of $E(\alpha)$, without specifying particular cases. 

First of all, we recall that normalizable bound-state solutions (for which $\beta > 0$) 
occur only for $|m_1-m_2| < E < m_1+m_2$.  There are two branches to the solution $E(\alpha)$, 
much like for the scalar Yukawa model \re{1.20} described in the introduction.  The upper branch,
$E_+(\alpha)$, begins with the Balmer form $E_+ = m_1 + m_2 -
 {1 \over 2} {{m_1 m_2} \over {m_1 + m_2}}{{ \alpha^2} \over {n^2}} + 
O(\alpha^4)$ at low $\alpha$, then decreases monotonically towards a limiting value $E_c$.
There is also a lower `unphysical' branch that starts off 
 as $E_- = |m_1-m_2| +  {1 \over 2} {{m_1 m_2} \over {|m_1 - m_2|}}{{ \alpha^2} \over {n^2}} 
+ O(\alpha^4)$ for $m_1 \ne m_2$ but as $E_- = m \alpha/n + O(\alpha^3)$ for 
 $m_1 = m_2 = m$,  and increases monotonically towards $E_c$. 
The lower branch is not of Balmer form at low $\alpha$.  

The qualitative behaviour of $E(\alpha)$ is different for the fermion-like case $m_1 < m_2$ and 
the scalar-like case $m_1 > m_2$. This is evident from the one-body limits, described below. 
For the fermion-like case there are bound state solutions  for all values of $\alpha$,
 no matter how large. Indeed, 
 as $\alpha \to \infty$, the upper branch $E_+ (\alpha)$ approaches the value 
$E_c = \sqrt{m_2^2 - m_1^2}$ from above, 
while the lower branch $E_- (\alpha)$ approaches this value from below. Thus, 
$E=E_c = \sqrt{m_2^2 - m_1^2}$ is a horizontal asymptote of $E(\alpha)$ for the 
fermion-like cases.

In contrast, for scalar-like ($m_1 / m_2 > 1$) cases, there are bound state solutions 
only for finite $\alpha \le 
\alpha_c$, beyond which $E$ ceases to be real.   The 
qualitative shape of $\alpha(E)$, is that of a distorted upper half-circle ($\alpha_c$ being 
the apex), reminiscent of the scalar Yukawa result \re{1.20}.
The critical point, $E_c(\alpha_c)$ is the 
end-point for both branches for the scalar-like case. The critical value of $\alpha$ 
varies with $m_1/ m_2 > 0$. We find that 
 $  \alpha_c / n > 1 $ for all scalar-like cases.  The value  $\alpha_c / n = 1$ corresponds to the one-body 
Klein-Gordon limit ($m_1/ m_2 \to \infty$) for which $E_c-m_1 = 0$.  
For $n' = 0$ states, $\alpha_c$ lies in the domain $1 \le \alpha_c / n \le 2$, where $\alpha_c/ n=1$ 
 corresponds to the equal-mass limit, $m_2=m_1$ (cf. Eq. \re{9.30}).  
For $n'>0$ states, $\alpha_c/n$ generally increases with increasing $n'$, and becomes 
arbitrarily large in the equal mass limit ($m_1/m_2 \to 1$). 
	We shall now discuss the limiting cases of the physical, upper branch of $E(\alpha)$ 
 in some detail. 
\vss
\ni \underline { One mass is large and the other is small }\\

First, we consider the one-body limits of $E_+ (\alpha)$, which follow from \re{9.1} by 
making the substitutions 
 $E=m_a
(\varepsilon +1)$, where $ \varepsilon= c_1 \frac{m_b}{m_a} +c_2  \left(\frac{m_b}{m_a}\right)^2 +....$,
 and solving for the coefficients $c_i$ . 
For  $m_1/m_2 <1$ we obtain the fermion-like or Dirac limit:
\begin{eqnarray}\label{9.2}
E_+ \left(\frac {m_1}{m_2}<1\right) = m_2+m_1\left[  
\sqrt{1-\frac{\alpha ^2}{(n'+\sqrt{\alpha^2+\kappa ^2 })^2} }
\right.\qquad\qquad\qquad\qquad\qquad\nn\\
\left. ~~~~ + ~
\frac{m_1}{m_2}
\frac{\alpha ^2}{2(n'+\sqrt{\alpha^2+\kappa ^2 })^2}
\left(
1 - \frac{2\alpha ^2}{\sqrt{\alpha^2+\kappa ^2 }(n' + \sqrt{\alpha^2+\kappa ^2 })}
\right) + O \left( \left(\frac {m_1}{m_2}\right)^2 \right)   \right]
\end{eqnarray}
The first term in the square brackets is just the  Dirac  one body energy spectrum for a 
Coulombic potential (scalar coupling), and the second one
gives the first order correction in $m_1/m_2$.

For $m_2 / m_1 < 1$ the expansion yields the scalar or Klein-Gordon limit:
\begin{equation}\label{9.3}
E_+ \left(\frac {m_2}{m_1}< 1\right) =
m_1+m_2\left[  \sqrt{1-\frac{\alpha ^2}{n^2} }
+\frac{m_2}{m_1}\frac{\alpha ^2}{2n^2} + O \left( \left(\frac {m_2}{m_1}\right)^2 \right)  
\right]
\end{equation}
The first term in square brackets coincides with the Klein-Gordon one body
 energy spectrum in a Coulombic potential
 (scalar coupling), while the second gives the first order  correction to it, in powers of $m_2/m_1$.

\vss
\ni \underline { Expansion of $E$ in powers of the coupling constant }\\

Next, we consider the 
expansion of $E_+ (\alpha)$ in powers of $\alpha$. The result is 
\begin{equation}\label{9.4}
E_+ (\alpha) =  m_1 + m_2 - \frac{1}{2} m_r \frac {\alpha ^2}{n^2} -
\frac{1}{8} m_r  \alpha ^4  
\left[ \left(1 + \frac{m_1 m_2}{(m_1+m_2)^2} \right)
 \frac{1}{n^4} -    4 \frac{m_2^2}{(m_1+m_2)^2} \frac{1}{n^3 (j+ \frac{1}{2})}  
                  \right] + O(\alpha^6) ,
\end{equation}
where $\ds n = n' + j + \frac{1}{2}$ and  $\ds m_r = \frac{m_1 m_2}{m_1 + m_2}$.
We obtain the expected non-relativistic Balmer result at O($\alpha^2$). The O($\alpha^4$) 
correction is not symmetric in $m_1$ and $m_2$ due to the different, fermionic and bosonic, 
nature of the particles. The one body limits of \re{9.4} have the required Dirac and 
Klein-Gordon forms, as can be seen from the expressions
\begin{eqnarray}\label{9.5}
E_+\left(\frac{m_1}{m_2} < 1\right)  =  m_2+m_1\left[1
-\frac{\alpha ^2}{2n^2}-
 \frac{1}{8} \alpha ^4 \left(\frac{1}{n^4}-   \frac{4}{n^3 (j+ \frac{1}{2}) }   \right) 
\right. \qquad\qquad\nn\\  
 \left. +~
\frac{m_1}{m_2}\left( \frac{\alpha ^2}{2n^2} -   \frac{3 \alpha^4 }{2n^3(j+ \frac{1}{2})}   \right)
 +  O\left( \left(\frac {m_1}{m_2}\right)^2\!,~\alpha^6 \right)
\right],
\end{eqnarray}
\begin{equation}\label{9.6}
E_+ \left(\frac{m_2}{m_1} < 1\right) =  m_1+m_2\left[1
-\frac{\alpha ^2}{2n^2}-
\frac{\alpha ^4}{8n^4}+\frac{m_2}{m_1}\frac{\alpha ^2}{2n^2}+ 
O \left( \left(\frac {m_2}{m_1}\right)^2\!,\alpha^6 \right)
\right].
\end{equation} 
Note that \re{9.5} and \re{9.6} are the same as the expansions of 
 Eqs. \re{9.2}, \re{9.3} in powers of $\alpha$, as they ought to be.


\section{Concluding remarks}
\vss
We have studied two-particle systems in a model QFT, in which fermions of mass $m_1$ interact
 with bosons of mass $m_2$. The interaction is mediated by a real, scalar field of mass $\mu$ (the 
`chion' field).  The field equations were used to recast the Hamiltonian of the theory into a form in 
which the chion propagator appears directly in the interaction term.  For the case where there is 
no decay, emission or absorption of real (physical) chions ({\sl i.e.} only 'virtual' chions), we 
obtain exact two-particle eigenstates of the Hamiltonian, using an unconventional `empty' vacuum 
state, which is annihilated by both the positive and negative frequency parts of the particle-field 
operators.  

The resulting relativistic two-particle wave equation, for the stationary states of the system, 
reduces to a pair of Dirac-like radial equations for the various $J^P$ states.  These equations 
are shown to have the radial Schr\"odinger equation for the relative motion of the two particles 
as the non-relativistic limit, and the Dirac and Klein-Gordon equations (with scalar coupling) as the 
one-body limits.  Analytic solutions for the two-body bound state eigenenergies (rest masses) 
are obtained for the massless chion exchange ($\mu=0$) case.  The shape of the $E(\alpha)$, or 
$\alpha(E)$, curves, where $\alpha$ is the dimensionless coupling constant, is discussed for 
various mass combinations, $m_1/m_2$, and various $nJ^P$ states.  

In the case of massive chion exchange ($\mu \ne 0$), the eigenvalues and eigenfunctions must be 
obtained numerically, which can be done easily by standard methods.  We do not 
present such solutions in this paper. Also, we do not discuss the 
scattering-state solutions of the equations, though these can  be worked out readily.

Lastly, we mention that $\cal N$-body eigenstates, where ${\cal N} \ge 3$,
and the corresponding 
relativistic $\cal N$-body equations, can be worked out readily for the
present model, as was 
shown for the purely scalar model in \cite{Dar98}, and for QED in
\cite{DKluw}. Such equations 
are much more complicated, since they possess all the complexity of
relativistic many-body 
equations. We do not discuss the ${\cal N} \ge 3$ systems in this paper.






\end{document}